\documentclass[pra,superscriptaddress,a4paper,twocolumn,floatfix]{revtex4}

\usepackage{graphics} 
\usepackage{epsfig} 

\begin{document}   

\title{Strongly Enhanced Thermal Stability \\ of Crystalline Organic Thin
  Films \\ Induced by Aluminum Oxide Capping Layers}

\author{Stefan Sellner}

\affiliation{Max-Planck-Institut f\"ur Metallforschung, Heisenbergstr. 3,
  70569 Stuttgart, Germany}
\affiliation{Institut f\"ur Theoretische und Angewandte Physik, Universit\"at
  Stuttgart, Pfaffenwaldring 57, 70550 Stuttgart, Germany}

\author{Alexander Gerlach} 
\author{Frank Schreiber}
\email[corresponding author:]{frank.schreiber@chem.ox.ac.uk}

\affiliation{Physical and Theoretical Chemistry Laboratory, Oxford University,
  South Parks Road, OX1 3QZ, United Kingdom} 

\author{Marion Kelsch} 

\affiliation{Max-Planck-Institut f\"ur Metallforschung, Heisenbergstr. 3,
  70569 Stuttgart, Germany}

\author{Nikolai  Kasper} 

\affiliation{Max-Planck-Institut f\"ur Metallforschung, Heisenbergstr. 3,
  70569 Stuttgart, Germany}
\affiliation{ANKA, FZ Karlsruhe, Hermann-von-Helmholtz Platz 1,
  76344 Eggenstein-Leopoldshafen}

\author{Helmut Dosch}

\affiliation{Max-Planck-Institut f\"ur Metallforschung, Heisenbergstr. 3,
  70569 Stuttgart, Germany}
\affiliation{Institut f\"ur Theoretische und Angewandte Physik, Universit\"at
  Stuttgart, Pfaffenwaldring 57, 70550 Stuttgart, Germany}

\author{Stephan Meyer}
\author{Jens Pflaum}

\affiliation{III. Physikalisches Institut, Universit\"at Stuttgart,
  Pfaffenwaldring 57, 70550 Stuttgart, Germany}

\author{Matthias Fischer}
\author{Bruno Gompf}

\affiliation{I. Physikalisches Institut, Universit\"at Stuttgart,
  Pfaffenwaldring 57, 70550 Stuttgart, Germany}

\begin{abstract}
  We show that the thermal stability of thin films of the organic
  semiconductor diindenoperylene (DIP) can be strongly enhanced by aluminum
  oxide capping layers. By thermal desorption spectroscopy and
  \textit{in-situ} X-ray diffraction we demonstrate that organic films do not
  only stay on the substrate, but even remain crystalline up to 460$^\circ$C,
  i.e.\ 270$^\circ$C above their desorption point for uncapped films
  (190$^\circ$C). We argue that this strong enhancement of the thermal
  stability compared to uncapped and also metal-capped organic layers is
  related to the very weak diffusion of aluminum oxide and the structurally
  well-defined as-grown interfaces. We discuss possible mechanisms for the
  eventual breakdown at high temperatures.\\[1em]

\end{abstract}

\date{\today}

\maketitle

\section{Introduction}
\label{sec:intro}

Organic electronics is considered to be one of the key areas of future
thin-film-device technology. Several device applications have already been
shown to exhibit convincing performance, organic light-emitting diodes being
one of the most successful
examples\cite{forrest-chrev97,ho-nat00,dimitrakopoulos-am02}. However, besides
the obvious performance requirements, the devices have to meet stability
standards, which in some cases are actually the limiting factor of
technological progress\cite{farchioni-book}.  Indeed, stability at elevated
temperatures, high electrical field gradients, and against exposure to
corrosive gases like oxygen is crucial for all commercial applications.

It has turned out that thermal stability of thin organic films is not only
related to fabrication procedures, but constitutes rather fundamental
challenges\cite{schreiber-pss04}. It is thus a prerequisite to understand and
to control\cite{salaneck-book}:\\
-interdiffusion at organic/metal interfaces during and after growth\\
-thermally induced dewetting effects at organic-inorganic interfaces\\
-structural phase transformations of the organic material at
temperatures often not far from temperatures of operation\\
-the vapor pressure of low-weight organics at  elevated temperatures.\\
Moreover, interfaces of organic films are often chemically and structurally
heterogeneous, and their controlled preparation is
non-trivial\cite{peumans-nat03}. This applies to metallic
contacts\cite{duerr-jap03,duerr-adm02,koch-tsf03,faupel-matent98}, to
insulating layers\cite{lee-synmet03,ha-jncs02}, which are typically required
in field-effect geometries, as well as to organic-organic interfaces as found
in photovoltaic elements\cite{forrest-chrev97,fenter-chphyslet97}.

In this paper we show that aluminum oxide layers can be prepared on organic
semiconductor films of diindenoperylene (DIP) with well-defined interfaces
which render thermally very stable heterostructures. This finding could open
the way for organic thin film devices to operate at significantly higher
temperatures than hitherto assumed.

\section{Results}
\label{sec:results}

\subsection{As-grown structure}

\begin{figure}[htbp]
  \centering
    \includegraphics[width=8.5cm]{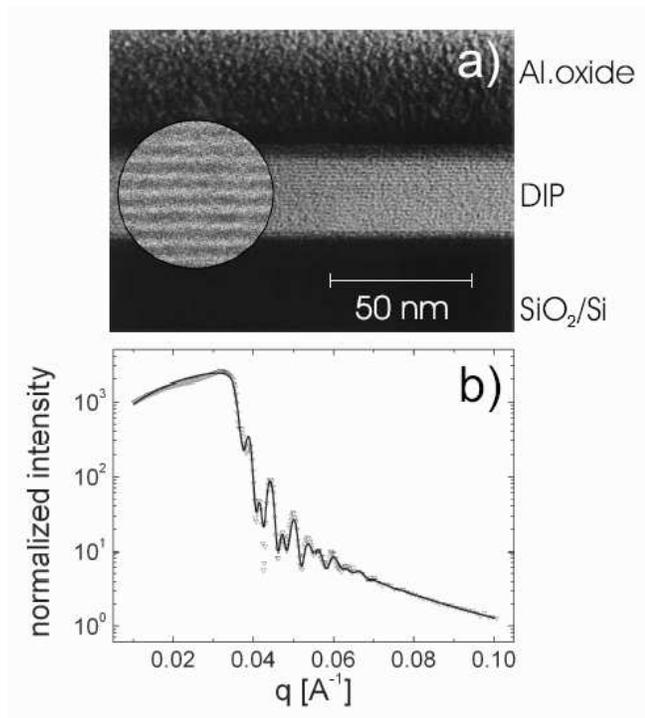}
    \caption{a) TEM image showing a well-defined heterostructure of aluminum
      oxide on DIP (300\,\AA) on silicon oxide together with the layer
      resolved structure of the organic film (inset). b) Room temperature
      X-ray reflectivity scan with least-square fit of an aluminum oxide
      capped DIP film showing pronounced Kiessig interference fringes.}
  \label{fig:tem}
\end{figure}

Thin films of DIP exhibit high structural
quality\cite{duerr-apl02,duerr-prl03}.  After depositing aluminum oxide on the
organic film the TEM image (Fig.~\ref{fig:tem}a) shows that the resulting
heterostructure is very well-defined. Laterally homogeneous interfaces and
only limited interdiffusion are observed. Within the organic film individual
molecular layers of DIP can be resolved as indicated by the inset.
The X-ray reflectivity data with pronounced Kiessig interferences -- taken on
a different sample -- confirm this picture (Fig.~\ref{fig:tem}b).  The Kiessig
fringes correspond to a film thickness of 573\,\AA\/ for the aluminum oxide and
1020\,\AA\/ for the DIP film. Laue oscillations around the Bragg peak (see
Fig.~\ref{fig:reflectivity}b below) of DIP confirm its high crystallinity,
with the coherent thickness equaling the total thickness, implying that the
DIP film is coherently ordered across its entire thickness.  The well-defined
character of the as-grown interfaces makes these heterostructures ideally
suitable for studies of the thermal stability.

\subsection{Temperature dependence: Thermal desorption data}

\begin{figure}[htbp]
  \centering
    \includegraphics[width=8.5cm]{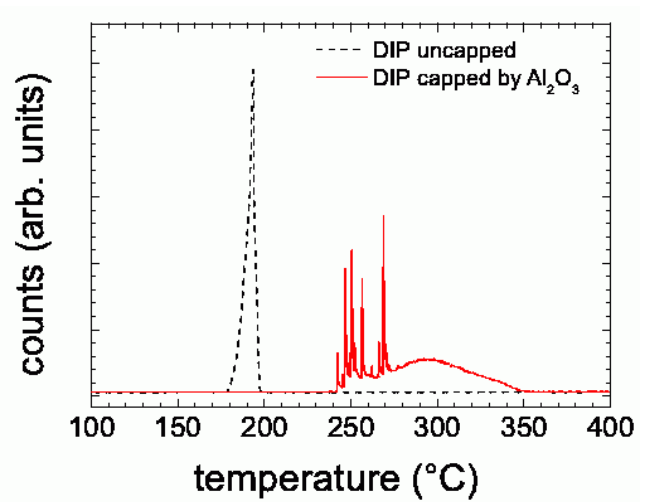}
  \caption{Thermal Desorption Spectra of an uncapped vs. a capped DIP
    film. The uncapped DIP film (broken line) shows only one well-defined peak
    at 190$^\circ$C. The desorption features for the capped DIP film (solid
    line) are clearly shifted to higher temperatures.}
  \label{fig:tds}
\end{figure}

Fig.~\ref{fig:tds} shows thermal desorption spectroscopy data which
demonstrate the enhancement of the thermal stability of capped DIP films
compared to uncapped layers. In these experiments the mass spectrometer is
tuned to the mass of DIP molecules (400 amu), and the signal is recorded as a
function of time, while the temperature is ramped at a constant rate of
0.5$^\circ$C/sec.  While the uncapped DIP shows a well-defined desorption peak
around 190$^\circ$C the capped film shows no desorption until 240$^\circ$C.
The shape of the spectra also reveals differences in the desorption
\textit{process}.  The spectrum of the uncapped film shows only one sharp peak
that can be attributed to the DIP 'bulk' desorption.  In contrast, the
desorption spectrum of the capped film extends over a broader temperature
range and has multiple features with the main feature centered at
300$^\circ$C.

We regard the sharp spikes as evidence for isolated desorption channels such
as small cracks, which do not give rise to desorption of the entire film.
Only a small (local) fraction of DIP molecules may desorb at higher rate and
on short time scales. The existence of these spikes was confirmed for several
samples, but their exact position and height depends on the individual sample,
consistent with the notion of the spikes being related to the properties of
the individual capping layer. The broad desorption feature extending to high
temperatures is related to the dominating part of the sample (the area under a
given feature is proportional to the number of desorbing molecules).  The wide
temperature range may be taken as an indication of a laterally inhomogeneous
capping layer and thus desorption processes from locally substantially
different regions.  Also, the state of the sample changes to some extent
during the heating process, and one may speculate that the thermally
stimulated desorption from a stable existing structure is also accompanied by
a certain rate of thermally induced formation of cracks, and thus an increase
of desorption from these defects which will depend on the exact heating rate.
This is consistent with our finding that with decreasing heating rate the
observed structural breakdown in TDS measurements shifts to higher
temperatures.

\subsection{Temperature dependence: X-ray data}

\begin{figure}[!ht]
  \centering
    \includegraphics[width=8.5cm]{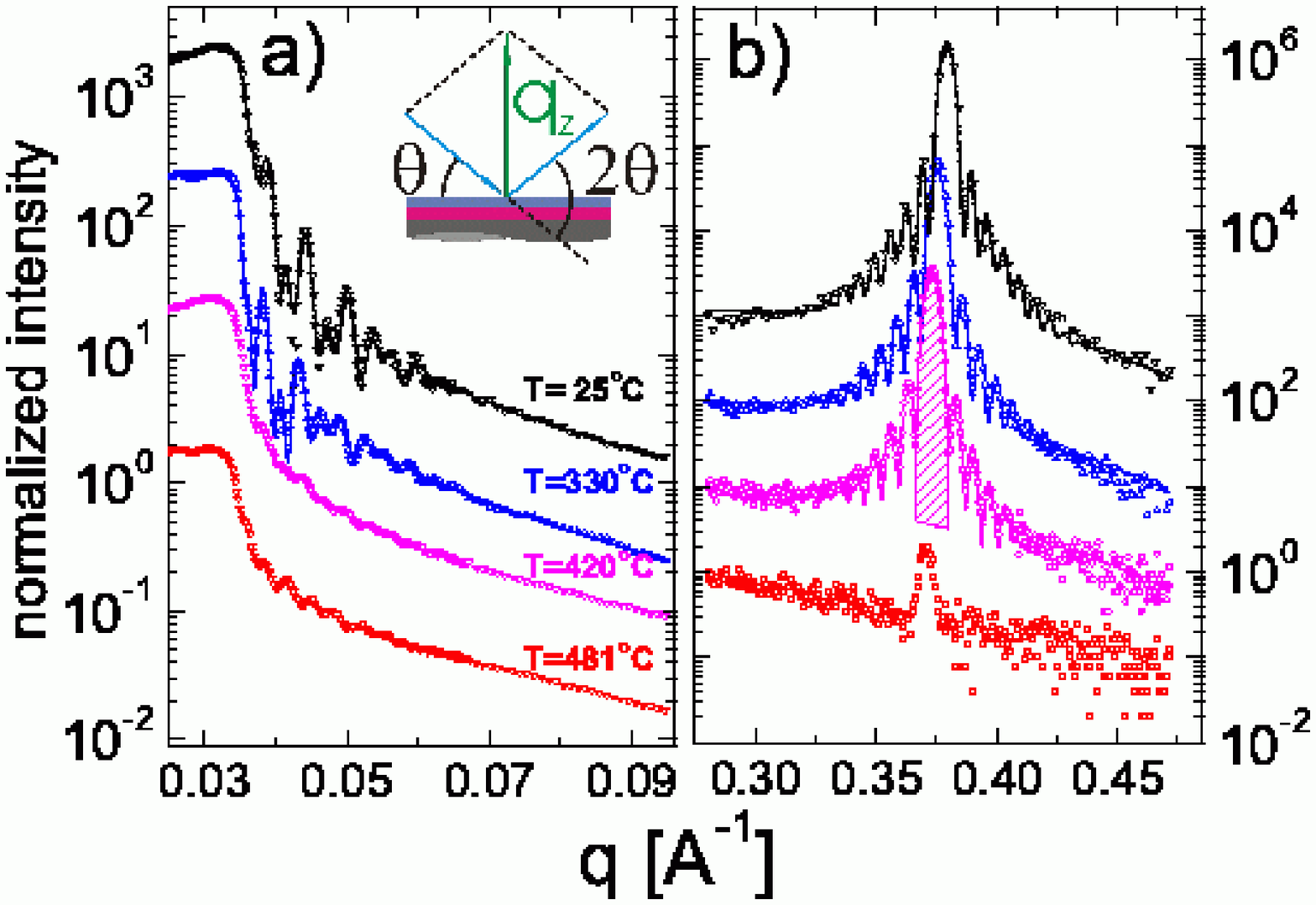}
  \vfill
    \includegraphics[width=8.5cm]{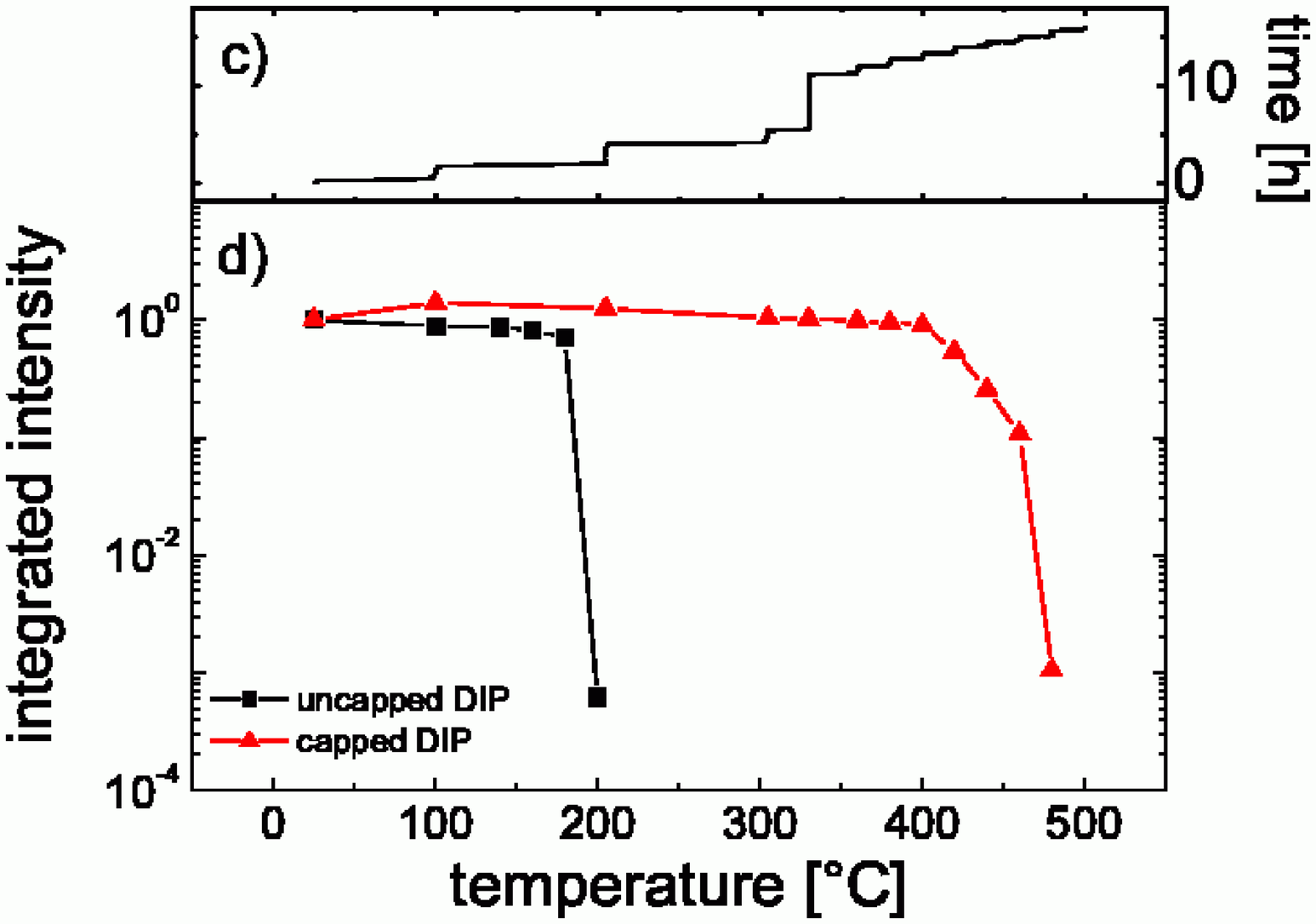}
  \caption{(Upper panel) X-ray reflectivity data of the aluminum
    oxide/DIP heterostructure with least-square fits at different
    temperatures.  The scattering geometry is indicated in the inset. By
    heating the sample the initially well-defined Kiessig fringes in panel~(a)
    slowly degrade.  The electron density profile can be reconstructed using
    the Parratt formalism\cite{tolanbook} after correcting for diffuse
    scattering. Most prominently the roughness of the DIP/aluminum oxide and
    aluminum oxide/vacuum interfaces increase with higher temperatures.  The
    first order Bragg peak with Laue oscillations in panel~(b) remains visible
    up to $T=460^\circ$C. For clarity the datasets are plotted with an offset.
    The comparison of the DIP film with and without aluminum oxide capping
    layer reveals the increased stability of the multilayer system as
    indicated by the integrated intensity of the Bragg peak as a function of
    temperature (d). For the temperature ramp in these experiments as
    displayed in panel (c) for the capped DIP film this results in a breakdown
    at $T=460^\circ$C compared to $T=190^\circ$C for uncapped DIP films.}
  \label{fig:reflectivity}
\end{figure}

In order to shed more light on the degradation process and the 'kinetics' of
the breakdown at high temperatures we performed \textit{in-situ} X-ray
diffraction experiments (Fig.~\ref{fig:reflectivity}).
The films are heated up from 25$^\circ$C to 500$^\circ$C in a stepwise fashion
and after thermal equilibration (on a time scale of a few minutes) X-ray
reflectivity scans (Fig.~\ref{fig:reflectivity}a and b) are taken at each
intermediate temperature. Since the film is kept at elevated temperatures for
several hours (see Fig.~\ref{fig:reflectivity}c) the corresponding averaged
'heating rate' of $\sim 0.01^\circ$C/sec is of course much lower than in the
TDS experiment.

As can be seen from Fig.~\ref{fig:reflectivity}a, Kiessig interference fringes
are clearly visible up to $\sim 380^\circ$C, but are gradually damped out for
still higher temperatures.  Importantly, the Bragg reflection at $q \approx
0.38$\,\AA$^{-1}$ (Fig.~\ref{fig:reflectivity}b) remains virtually unchanged
up to $440^\circ$C, showing that the DIP crystal structure stays intact. Until
up to 460$^\circ$C, the Laue oscillations around the DIP Bragg peak show no
significant changes implying that the organic film does not only remain
crystalline, but also coherently ordered.

Fig.~\ref{fig:reflectivity}d compares the integrated intensity of the DIP
Bragg peak with and without aluminum oxide capping layer for the given
temperature-time protocol (Fig.~\ref{fig:reflectivity}c).  As reported in
previous studies\cite{duerr-jap03,duerr-adm02} and in agreement with TDS,
uncapped DIP films desorb already at about $T=190^\circ$C.  In contrast, films
with an aluminum oxide capping layer can be stabilized up to $T=460^\circ$C,
i.e.\ well beyond the desorption temperature of the uncapped DIP layer.

\begin{figure}[!ht]
  \centering 
    \includegraphics[width=8.5cm]{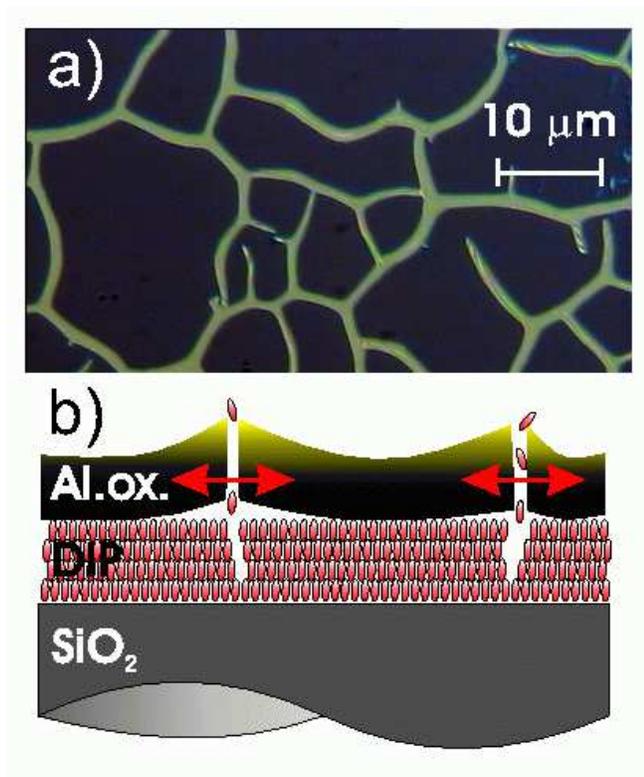}
  \caption{a) Optical micrograph of the surface of the sample after heating.
    We interprete the bright lines as fractures in the capping layer induced
    by the thermal strain upon heating. b) Sketch of the degradation scenario
    resuming the experimental results.}
  \label{fig:model}
\end{figure}

In a separate set of experiments, in order to evaluate long-term effects of
heating, one reference sample was kept at $300^\circ$C and the DIP Bragg peak
with Laue oscillations was recorded repeatedly for more than 300 hours. Its
integrated intensity is found to decrease with time 
-- but exhibits still $\ge$  50$\%$
of its initial value after $\sim 130$ hours -- 
whereas the coherent thickness
of the organic film remains unchanged over the entire period.  This suggests
that the observed degradation process of the capped DIP film is kinetically
limited by desorption from defects within the capping layer. Since the
coherent thickness stays constant the decrease of the integrated intensity at
a certain temperature with time is related to desorption of the organic film
from areas near thermally induced defects in the capping layer, as e.g.\ 
microcracks or holes.  In a second step, even molecules from well-capped
domains diffuse to these defects as a function of temperature and time
(Fig.~\ref{fig:model}).

\section{Discussion and Summary}
\label{sec:discuss}

While an ideal capping layer is of course \textit{expected} to suppress
evaporation of the organic layer underneath, the remarkable finding is that,
given the inevitable defects of real samples, the capping not only does
enhance the stability, but does it so effectively.  The increase of the
thermal stability by up to $270^\circ$C in our experiments has to be seen in
comparison with, e.g., metal capping layers which interdiffuse at low
temperatures and tend to compromise the organic layer already upon deposition.
We note that for accidentally non-stoichiometric aluminum oxide layers,
specifically those with higher metal content, the DIP film structure broke
down at temperatures in between $190^\circ$C and $450^\circ$C.  We take this
as an indication of the metal atoms with their higher mobility being
responsible for the weaker stabilization effect.

The oxide capping layer apparently is relatively near to the concept of a
'closed' layer and does not interdiffuse strongly upon deposition. Equally
important, the amorphous aluminum oxide is less prone to diffusion than gold,
so that even at elevated temperatures the capping layer presumably does not
'move' much, in contrast to gold, which has a significant mobility at the
temperatures relevant in this study.  Nevertheless, also the oxide-capped DIP
films ultimately break down.  The detailed scenario of this process is still
under investigation, but we speculate that it is related to minor cracks or
narrow channels within the aluminum oxide (Fig.~\ref{fig:model}a and b).  Also
chemical decomposition of DIP at high temperatures may play a role.  Since the
multilayer structure obviously depends strongly on the preparation conditions,
the enhancement of the thermal stability will also vary accordingly. 

The evidence of the enhanced thermal stability is of great practical
importance, not only for the specific system of DIP studied here, but for
organic electronics in general.  It offers a route for the stabilization of
compounds with vapor pressures so far considered too high for utilization in
organic-based devices, thus extending the range of applications and working
conditions, including harsher environments and elevated operating temperatures
up to the thermal degradation of the molecules. The electrical
characterization of capped semiconductors is presently underway.

\section*{Experimental}
\label{sec:exp}

Thin films of DIP (thickness 300--1000\,\AA) on Si(100) wafers with native
oxide were prepared by organic molecular beam deposition (rate 12\,\AA/min,
growth temperature 145$^\circ$C)\cite{duerr-jap03,duerr-apl02}.  The aluminum
oxide layers were deposited on DIP by HF-magnetron sputtering in a separate
chamber.  To avoid oxidation of the underlying organic film we use pure argon
as sputter gas (argon partial pressure $3\times 10^{-3}$\,mbar). After some
sputter cycles this leads to an under-stoichiometric target with respect to
oxygen content.  To overcome this problem the target was refreshed after each
deposition in an oxygen/argon atmosphere.  The base pressure of the chamber
was $3 \times 10^{-7}$\,mbar, the deposition rate about 7\,\AA/min and the
substrate temperature was kept at $-10^\circ$C.  Aluminum oxide films
deposited under these conditions are totally transparent and amorphous.

The samples have then been studied by transmission electron microscopy (TEM),
thermal desorption spectroscopy (TDS), and \textit{in-situ} X-ray reflectivity
at the Max-Planck beamline at the ANKA synchrotron source (FZ Karlsruhe,
Germany). The data were taken from the centre of the sample so that edge
effects can be excluded.

The temperature dependent X-ray reflectivity studies were done using a small
vacuum chamber with a capton window and an integrated sample heater. All
temperatures were measured with two calibrated C-type thermocouples close to
the sample.

\paragraph*{\textbf{Acknowledgments}}

We acknowledge support by the Deutsche Forschungsgemeinschaft (DFG) within the
Focus Program on organic field effect transistors and by the Engineering and
Physical Sciences Research Council \mbox{(EPSRC)}. We are grateful to the FZ
Karlsruhe and the ANKA management for their generous support.

\end{document}